\RequirePackage{fix-cm}
\documentclass[twocolumn]{svjour3}          
\smartqed  
\usepackage{graphicx}
\usepackage{amsmath}
\usepackage{float}
\journalname{Journal of Computational Electronics}
\begin{document}

\title{Higher harmonics and {\em ac} transport from time dependent density
  functional theory}


\titlerunning{Higher harmonics and {\em ac} transport from TD-DFT}        
\author{Christian Oppenl\"{a}nder \and Bj\"{o}rn Korff \and Thomas A. Niehaus
}


\institute{C. Oppenl\"{a}nder   \at Department of Theoretical Physics, University of Regensburg, 93040 Regensburg, Germany   
           \and
           B. Korff \at Bremen Center for Computational Materials Science,
Universt\"{a}t Bremen, Am Fallturm 1,
28359 Bremen, Germany 
           \and
            T.A. Niehaus  \at Department of Theoretical Physics,
            University of Regensburg, 93040 Regensburg, Germany \\
           \email{thomas.niehaus@ur.de}        
}

\date{Received: date / Accepted: date}

\maketitle

\begin{abstract}
We report on dynamical quantum transport simulations for realistic
molecular devices based on an approximate formulation of
time-dependent Density Functional Theory with open boundary
conditions. The method allows for the computation of various
properties of junctions that are driven by alternating bias
voltages. Besides the {\em ac} conductance for hexene 
connected to gold leads via thiol anchoring groups, we also
investigate higher harmonics in the current for
a benzenedithiol device. Comparison to a classical quasi-static model reveals
that quantum effects may become important already for small {\em ac} bias
and that the full dynamical simulations exhibit a much lower number of
higher harmonics. Current rectification is also briefly discussed.       

\end{abstract}

\section{Introduction}
\label{intro}
Molecular electronics, the use of single molecules as functional
entities in nanoscale electronic circuits, received significant
interest over the past years \cite{Cuevas2010}. The synergy of
experimental achievements and first principles atomistic modeling was
especially fruitful in this field. Key to a correct interpretation of
conduction measurements in break junctions or scanning tunneling
microscopy (STM) is for example the precise knowledge of the molecular
geometry within the junction. This information can nowadays be
obtained from Density Functional Theory (DFT) in a routine
fashion. Besides structural data, DFT may also be used to estimate I-V
characteristics in conjunction with the Landauer-B\"{u}ttiker formalism
\cite{Brandbyge2002,Rocha2006,Pecchia2008}.

 With the enormous chemical diversity of even small to medium sized
molecules, the initial hope was to choose appropriate compounds for a
given desired circuit element and fine-tune their electronic structure
by suitable functionalization. While this approach met practical
problems \cite{McCreery2013}, several groups are now investigating the
possibility of using time-dependent fields to gain additional control
over the conduction process
\cite{Duli'c2003,Molen2008,Battacharyya2011}. As an example,
application of visible and UV light induces a reversible
ring-opening/ring-closure reaction in some organic molecules that
leads to sizable on/off ratios of the current
\cite{Molen2008,Tam2011,Kim2012}. Light emission from molecular
junctions may also provide further means to characterize the device
\cite{Qiu2003,Wakayama2004,Dong2004}. In many of these experiments,
the frequency of the the applied radiation is small compared to the
plasmon frequency of the metallic leads. One can then assume that the
effect of the time-dependent field is just a rigid shift of the contact
energy levels, leading to a sinusoidal bias potential across the
junction. The resulting picture of {\em ac} conduction raises a couple of
interesting questions in the context of molecular electronics. Do
conventional organic molecules behave as classical capacitors or
inductors, or does their {\em ac} response depend on frequency and chemistry
in a non-trivial way? This question was already addressed by Fu and
Dudley in the 1990s for the model system of a resonant tunneling diode
\cite{Fu1993}, and recently, several groups presented DFT based
calculations of {\em ac} conductances for realistic molecular junctions
\cite{Yu2007a,Yam2008,Yamamoto2010,Sasaoka2011,Hirai2011}.

The time-dependence of the voltage has also other measurable
consequences: If the I-V characteristic of a device is non-linear, the
Fourier transform of the time-dependent current will feature peaks at
multiples of the driving frequency $\omega_0$. These higher harmonics
were utilized in STM to characterize also non-conducting samples
\cite{Kochanski1989,Stranick1994}. With the currently available
instrumentation, harmonics at optical frequencies are clearly difficult
to detect. However, the generation of higher harmonics implies current
rectification (the change of {\em dc} current due to an additional {\em ac} bias)
and is therefore indirectly accessible. In recent experiments the
measured non-linearities and the rectified current were used to
estimate the field enhancement in metallic nanoconstrictions subjected
to visible light \cite{Ward2010,Arielly2011}. On the theoretical side, the modeling of
photocurrents was pioneered by Tien and Gordon \cite{Tien1963}, while Cuevas and
coworkers presented a number of detailed DFT based calculations on this
topic in recent years \cite{Viljas2007,Viljas2007a,Viljas2008}. A
direct estimation of the amplitude and number of higher harmonics in
realistic molecular junctions seems to be missing. In principle,
Green's function approaches in the energy domain can be utilized for
this purpose, but require truncations in the number of photons
absorbed and emitted in the tunneling process
\cite{Brandes1997,Platero2004,Cuevas2010}.
 
 In this article we apply an approximate method based on
time-dependent density functional theory to investigate the {\em ac}
response of several realistic molecular devices. The density matrix of
the device is propagated according to a Liouville-van Neumann equation
including the effects of dissipation and decoherence due to the
macroscopic leads. The latter are characterized by an embedding
self-energy in the wide-band approximation. At each time step the
potential in the device region is determined by solving the
Poisson equation and used to update the time-dependent DFT Hamiltonian
in a selfconsistent fashion. The method allows one to compute the
time-dependent current $I(t)$ for arbitrary time-dependent bias
potentials. After a short description of the numerical scheme in
Sec. \ref{theo}, we investigate the frequency dependent admittance
$G(\omega)$ for a typical molecular junction in Sec.~\ref{adm}. 
The remainder of the paper
(Sec. \ref{hh}) is devoted to an analysis of the higher harmonics in a
molecular junction based on benzenediol. The goal is here to contrast the
results from a fully quantum-mechanical calculation with classical
estimates.

\section{Method}
\label{theo}
\begin{figure}
  \includegraphics[scale=0.40]{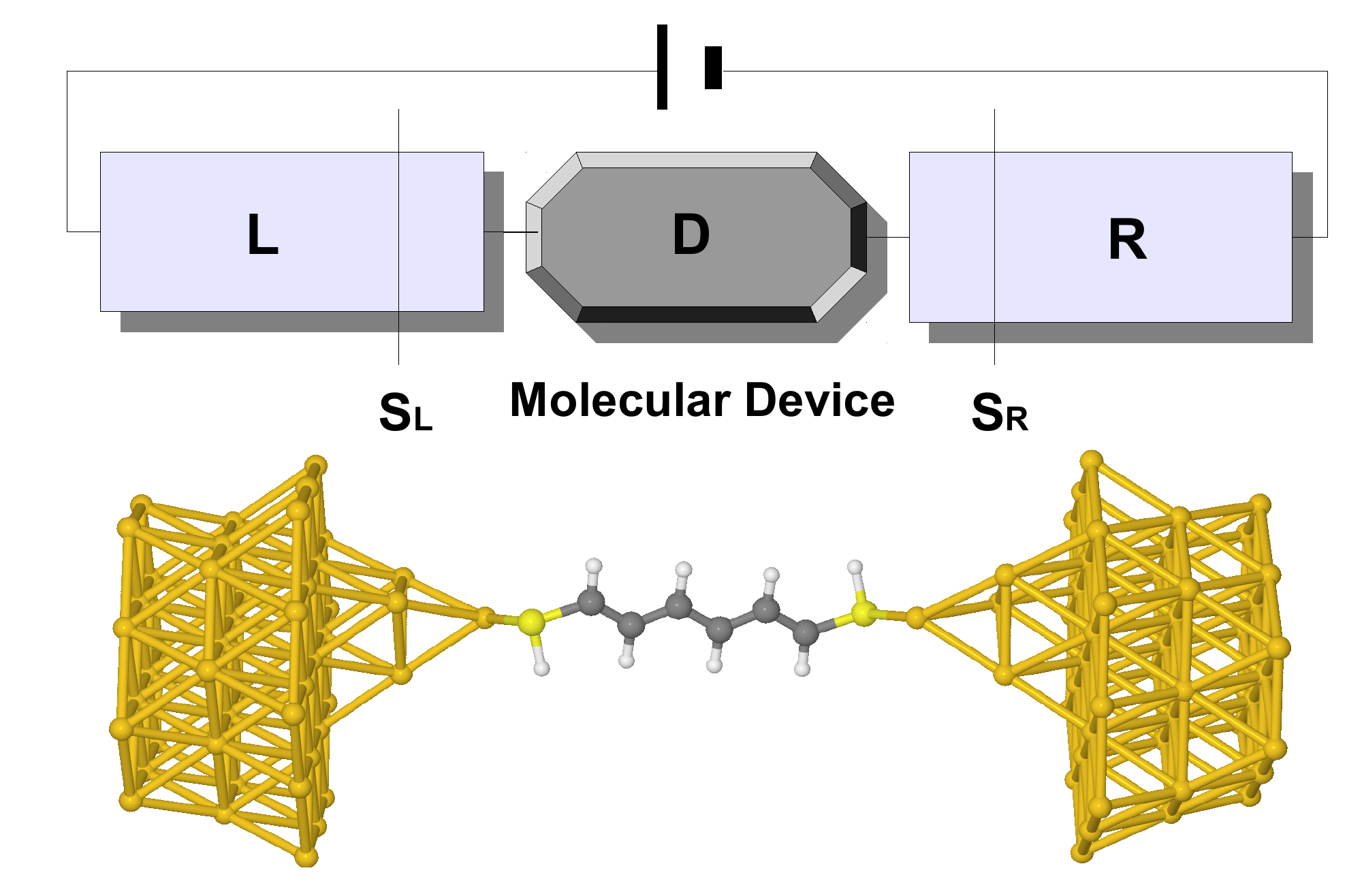}\\
  \caption{Top: Schematical representation of our theoretical setup. Bottom: The Au-Hexen-Au system. Only the device region is shown, the leads are reproduced periodically in either direction.}
  \label{goldsystem}
\end{figure}  

The following will outline the approach leading to the presented
results. Apart from the Hamiltonian entering the scheme, we follow the
method introduced by Zheng and co-workers \cite{Zheng2007}.
Our systems are partitioned into a central device region (D) - which also includes two layers of contact material - and semi-infinite periodic leads on either side of the device (L,R). The
leads are identical and the whole device is assumed to be in thermal
equilibrium ($ T = 0$ K) at a common chemical potential for $t=0$. In consequence, current only flows due to an applied time-dependent bias $V(t)$ of arbitrary shape and magnitude. As already
mentioned in the introduction, we employ a propagation scheme based on
the Liouville-van Neumann equation for the Kohn-Sham
density matrix. According to Ref.~\cite{Zheng2007}, a closed equation
for the reduced density matrix of the device region may be derived,
featuring dissipation terms 
 $Q_{\alpha=L/R}(t)$ arising from the interaction with the semi-infinite leads (in atomic units):

\begin{equation}
\label{neumann}
i\frac{\partial}{\partial t}\sigma(t) = [H(t),\sigma(t)] -i \sum_{\alpha=L,R} Q_\alpha(t).
\end{equation}

Here, ${\sigma(t)}$ is the reduced Kohn-Sham density matrix written in a basis of localized atom-centered basis functions $\phi_\mu(\mathbf{r})$. The $Q_{\alpha}(t)$ have been
derived in the framework of quantum dissipation theory using the
Keldysh non-equilibrium Green's function formalism.
It is possible to write down a formally exact and explicit expression for the dissipation terms using hierarchical equations of motions \cite{Zheng2010} applicable to arbitrary non-Markovian dissipative
systems. However, we aim for numerical efficiency and use the Markovian approximation to $Q_{\alpha}(t)$ proposed in the original article \cite{Zheng2007} within the wide band limit (WBL):

\begin{equation}
Q_{\alpha}(t)=i[\Lambda_{\alpha},\sigma(t)]+\{\Gamma_{\alpha},\sigma(t)\}+K_{\alpha}(t),\label{WBLQ}
\end{equation}

where $\Lambda_{\alpha}$ denotes the real part of the self-energy and $\Gamma_{\alpha}$ its imaginary part. The self-energy itself is calculated from the device-lead couplings and the surface
Green's function of the leads. The terms $K_{\alpha}(t)$ can also be readily
calculated from the Hamiltonian and the  applied bias potential. The
initial density matrix is obtained from equilibrium Green's function theory
using the same input Hamiltonian and the wide band limit for consistency. The equation of motion is then numerically propagated by a Runge-Kutta method. \\
In each time step, the time-dependent current can then be calculated through

\begin{equation}
 I_{\alpha}(t)=-\text{Tr}[Q_{\alpha}(t)].\label{current}
\end{equation}  

In the original scheme, the Hamiltonian governing the propagation is set to be from time-dependent density functional theory (TD-DFT \cite{Casida1995}) in the adiabatic approximation. However,
the systems we investigate involve more than a hundred heavy atoms in the device region, and time step converged trajectories require a resolution of ten attoseconds or less. At present state, it is
numerically extremely expensive to extract information about the
admittance behaviour for such demanding systems by use of
TD-DFT. Consequently, we employ the above propagation scheme in
conjunction with an approximate form of
TD-DFT, namely time dependent density functional based tight-binding (TD-DFTB) \cite{Frauenheim2002,Niehaus2009,Wang2011}  in addition to the WBL already mentioned. \\
The Hamiltonian matrix then takes the following form:

\begin{eqnarray}
\label{DeltaV}
H_{\mu\nu}(t) &=& \langle \phi_\mu | H[\rho_0]|\phi_\nu \rangle
\\\nonumber &+& \frac{1}{2}[\delta V_A(t)+\delta
V_B(t)]S_{\mu\nu},\quad \mu\in A, \nu \in B.
\end{eqnarray}  

Eq.~\ref{DeltaV} results from an expansion of the electron density in
the device region $\rho(\mathbf{r},t)$ around a constant reference
density $\rho_0$, assumed to be a superposition of the atomic densities $\rho_0=\sum_A \rho_A$. The zeroth order
represents a DFT Hamiltonian (with the PBE XC-functional
\cite{perdew1996gga} entering in our case) evaluated at the reference
density in a two-center approximation. Since the evaluation of this term only requires
information about the distance between atoms, it can be built from
pre-calculated lookup tables almost instantaneously. The second term
accounts for deviations from the reference density. At each time step, the charge density in the device region is computed from the
density matrix $\sigma(t)$ and the basis function overlap
$S_{\mu\nu}$. The solution of the Poisson equation with boundary
conditions matching the applied bias then determines the potential
shifts $\delta V_A(t)$. A more detailed description of the method may be
found in Ref.~\cite{Wang2011}.

\section{Admittance of a Au-Hexen-Au nanojunction}
\label{adm}

\begin{figure*}
   \includegraphics[scale=0.58]{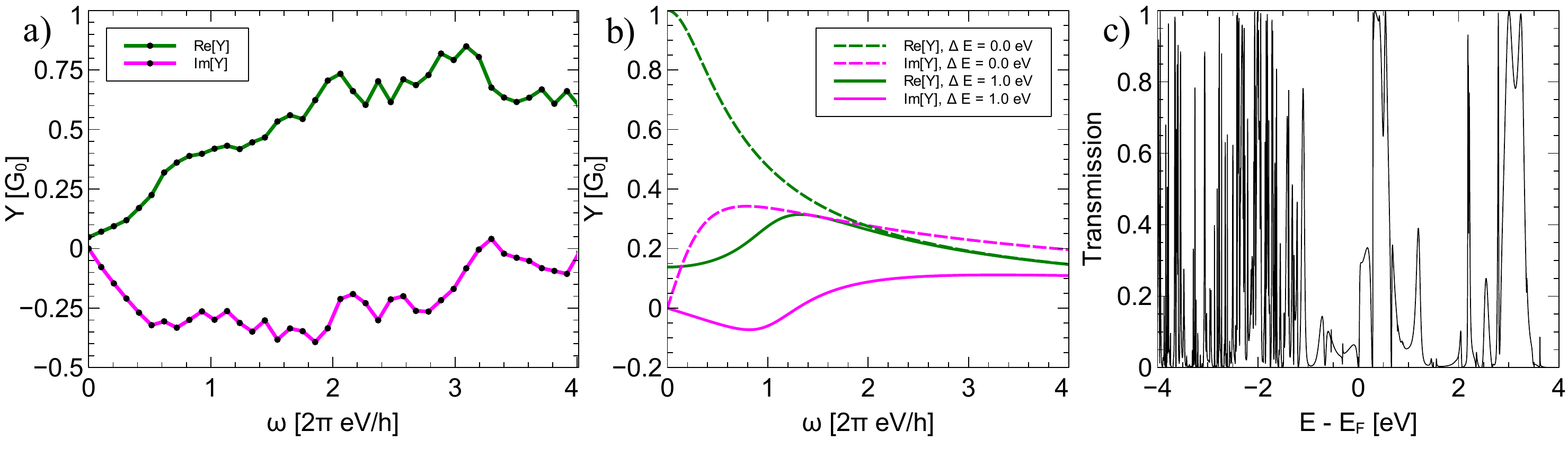}\\
   \caption{a) Conductance and susceptance of the
Au-Hexen-Au system. The current was evaluated for a total evolution time of $40$ fs using a
time step of $4$ as. b) Analytical
admittance for a single-level model according to
Eqs. \ref{fududr} and \ref{fududi} with $\gamma =
0.4$ eV and $\Delta E$ as indicated in the legend.  c) DFTB transmission of the Au-Hexen-Au junction
in the Landauer-B\"{u}ttiker formalism. }
\label{Allthree}
 \end{figure*}

Using the above approximations, we are able to treat junctions of
considerable size and can move away from mere test systems. We
investigated the admittance behaviour of a Hexene molecule coupled to
Gold leads using thiol bridges.  The leads consist each of 172 Au atoms
which set up a bulk by periodical replication and are used to
calculate the self-energy. The experimental lattice constant of
approximately 4.08 \r{A} \cite{Dutta1963} has been used.  The device
region includes two layers of the described contacts which are aligned
along the (111)-surface in transport direction with 86 Gold atoms in
total. Pyramids consisting of four atoms have been placed on top of the surfaces with
the same interatomic distances as in the bulk. The molecule has
been optimized in vacuum using DFT with the PBE functional and a
6-31G* basis set. In order to get a good guess for the binding angle,
an S-Au group was previously attached to either side. This approximate
geometry was then placed symmetrically between the pyramids without
further optimizations (see Fig.~\ref{goldsystem}). In our transport
calculations, all atoms are described within a minimal basis set: l=0 for
H, up to l=1 for C and S, and up to l=2 for Au. 

In the time-dependent simulation, we applied a Lorent\-zian input signal

\begin{equation}
  \label{dirac}
  V(t)= \frac{V_0}{\pi} \frac{\Gamma}{(t-t_0)^2 + \Gamma^2},
\end{equation}

with $\Gamma = 0.2$ fs, $t_0 = 5$ fs, $V_0 = 5 \cdot 10^{-5}$ V and
used its Fourier transform as well as the Fourier transform of the
current response to obtain a complex admittance curve according to
$Y(\omega) = I(\omega)/V(\omega)$.  The result for the Au-Hexen-Au
junction is depicted in Fig.~\ref{Allthree}a). Qualitatively, one can identify
capacitive behaviour for small frequencies $\omega$  in analogy to a classical
RC-circuit (up to second order) \cite{Yu2007a}:

\begin{equation}
  Y^\text{RC}(\omega) = -i \omega C + \omega^2 C^2 R.
\end{equation}

In an earlier study, we observed the quadratic increase of the conductance (Re[Y]) and the linear
decrease of the susceptance (Im[Y]) also in a simpler
system with little transmission in the vicinity of the Fermi energy $E_F$ \cite{Oppenlander2013}. 

For a single level model, the
admittance may be determined analytically as shown by Fu and Dudley
\cite{Fu1993}. Given a {\em dc} Breit-Wigner transmission of the form 
\begin{equation}
  \label{breit}
  T(E) = \frac{\gamma^2}{(E-E_0)^2 + \gamma^2},
\end{equation}
where $E_0$ is the level energy, the {\em ac} admittance reads 

  \begin{multline}
\label{fududr}
      \text{Re}[Y(\omega)] = \\G_0 \frac{\gamma}{2\omega} \left[
    \arctan\left(\frac{\Delta E + \omega}{\gamma}\right) -  \arctan\left(\frac{\Delta E - \omega}{\gamma}\right)\right]
  \end{multline}

and 

  \begin{multline}
\label{fududi}
      \text{Im}[Y(\omega)] = \\G_0 \frac{\gamma}{4\omega} \ln\left(
        \frac{\left[ (\Delta E + \omega)^2 + \gamma^2 \right]
          \left[ (\Delta E - \omega)^2 + \gamma^2
          \right]}{\left[\Delta E^2  + \gamma^2\right]^2} \right),
  \end{multline}

where $\Delta E = E_F - E_0$ and $G_0$ denotes the quantum of conductance. In Fig.~\ref{Allthree}b), we plot a visualization of these
formulas for different model parameters. If the conduction is
off-resonant in the {\em dc} limit, a capacitive regime with negative
values of Im[Y] occurs at finite $\omega$, while resonant transmission
leads to inductive behaviour. Generally, a broader peak in the
transmission also
leads to a significantly greater effect on the shape of the
admittance, especially its real part. Resonances at energies far
from $E_F$ contribute less to the admittance. Consequently, one
expects to be able to relate the profile of the admittance to broad features in the
{\em dc} transmission, at least close to the Fermi energy. 

We therefore computed also the transmission for the Au-Hexen-Au
junction in the Landauer-B\"{u}ttiker framework at the DFTB level \cite{Pecchia2008}. In Fig.~\ref{Allthree}c), one observes a large number of peaks with a width close
to zero in the energy range [-4,-2] eV. These likely correspond to localized metal states
that do not couple to the left and right lead and therefore play no role
for the conductance. Channels with a significant width and amplitude
occur around $E - E_F =$ -2 eV, 0.5 eV and 3 eV. 
The corresponding admittance given in  Fig.~\ref{Allthree}a) shows some similarities with the
analytical model: 
below a resonance with $\hbar\omega = |E_0 - E_F|$, the conductance rises
and the susceptance becomes more negative, as
associated with a capacitive system. This can most clearly be seen
near the first resonance at $\hbar\omega = 0.5$ eV. Also at $\hbar\omega =$ 2 eV and 3 eV, we can see
further maxima in the real part of the admittance. A further
comparison of numerical and analytical results is difficult, as the
admittance of a device with multiple transmission channels will very likely differ from a
simple superposition of single-level admittance traces. We are
therefore currently exploring the extension of the Fu and Dudley
result to the multi-level case.

\section{Higher harmonics}
\label{hh}
Given a time-dependent bias of the form
\begin{equation}
  \label{td-bias}
  V(t) = V_\text{dc} +
V_\text{ac} \cos(\omega_0 t),
\end{equation}
one expects for non-linear devices a current response at
multiples of the driving frequency $\omega_0$. This can most easily be
seen if one assumes that the time-dependent current follows the bias
instantaneously and is well characterized by the current-voltage
characteristic $I_\text{dc}(V_\text{dc})$ in the absence of harmonic
driving. This allows one to write
\begin{eqnarray}
  \label{class}
  I(t) &\approx& I_\text{dc}(V_\text{dc} +
V_\text{ac} \cos(\omega_0 t)) \nonumber\\
      & = & I_\text{dc}(V_\text{dc}) + \sum_{n=1}^\infty \frac{1}{n!}
      \left.\frac{d^{n} I_\text{dc}}{d V^{n}}\right|_{V_\text{dc}} \left(V_\text{ac} \cos(\omega_0 t)\right)^n.
\end{eqnarray}
Taking the expansion up to second order one obtains
\begin{eqnarray}
   I(t) &\approx& I_\text{dc}(V_\text{dc}) + \left.\frac{dI_\text{dc}}{dV}\right|_{V_\text{dc}}
    V_\text{ac}  \cos(\omega_0 t)\nonumber\\
   &&   + \frac{1}{4} \left.\frac{d^2I_\text{dc}}{dV^2}\right|_{V_\text{dc}} V^2_\text{ac}
       \left[1- \cos(2\omega_0 t)\right].  \label{sec}
\end{eqnarray}

These expressions explicitly show that a non-linear I-V will lead to
higher harmonics in the current. Moreover, the time-averaged current $\overline{I(t)}$
will in general differ from the {\em dc} one, i.e., one observes current
rectification. In recent experiments on nanoscale devices, the harmonic
driving is realized by irradiation of the junction with frequencies in
the microwave or optical range. Since the current is determined by the
local voltage difference across the device, the {\em ac} bias entering
Eq. \ref{class} can not be obtained from the external field
alone. Instead, the screening properties of the leads have to be taken
into account, including especially also plasmonic excitations for
higher frequencies. Assuming the validity of Eq. \ref{sec}, this
drawback might be turned into an advantage. Measuring both the
photocurrent $I_\text{ph}=\bar{I}-I_\text{dc}$ and $d^2I/dV^2$, the local field
enhancement may be estimated, which is difficult to access otherwise \cite{Tu2006,Ward2008,Ward2010,Arielly2011}. 

Given this practical importance it is worthwhile to investigate under
which circumstances Eqs. \ref{class} and \ref{sec} are reasonable
approximations. For Eq. \ref{sec} to hold, $V_\text{ac}$ must be small
and $I_\text{dc}$ weakly non-linear around $V_\text{dc}$. For metallic
nanoconstrictions the latter criterion is fulfilled for a relatively
large bias range while for molecular junctions strong non-linearities
arise close to molecular resonances. Around zero bias, the criterion is therefore more
easily matched for fully saturated compounds compared to conjugated
molecules, because of the larger gap in the former case. 

The assumption on which the first line in Eq. \ref{class} is based
seems to be more severe. In a reactive circuit, current and voltage
will in general be out of phase and it is not at all obvious that the 
{\em dc} I-V curve is relevant for the transport at high frequencies. In
fact, the admittance calculations of the last section clearly showed
that in molecular junctions already small frequency driving leads to
sizable changes in the conductance. Moreover, Eq. \ref{class} neglects
quantum effects. In the Tien-Gordon picture for photo-assisted
tunneling \cite{Tien1963}, the electrons may absorb multiple quanta of the
radiation field in the leads before tunneling. As a consequence, the
steady state transmission is probed at energies $E+n\hbar
\omega_0$. Still, Eq. \ref{sec} might be a reasonable approximation in
the limit of small $V_\text{dc}$, given that the transmission is
weakly changing on the scale $\hbar
\omega_0$ \cite{Tucker1985,Cuevas2010}.                             

\begin{figure}[h]
  \includegraphics[height=3cm]{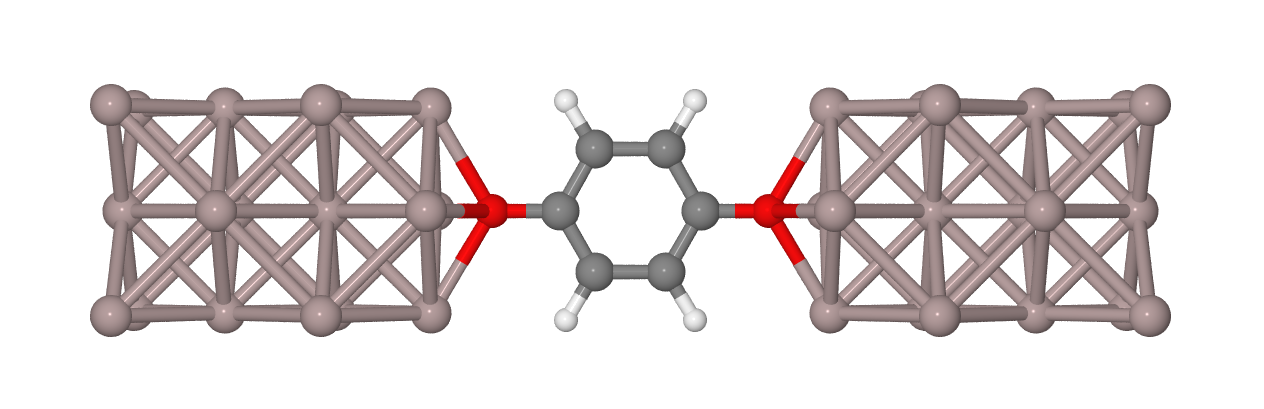}
  \caption{1,4-benzenediol inbetween Al nanowires of finite cross
sections oriented in (001) direction. The device region consists of
the molecule and 36 additional Al atoms, while the simulation cell for the
leads included 72 atoms.}\label{al-benze}
\end{figure}
In order to further investigate the limits of Eqs. \ref{class} and
\ref{sec}, we now present atomistic calculations on the device
depicted in Fig. \ref{al-benze}, a 1,4-benzenediol molecule coupled to an Al nanowire. This
junction was already characterized in earlier work
\cite{Yam2011,Wang2011} and the admittance
$G(\omega)$ has also been computed in Ref.~\cite{Oppenlander2013}. Here
we compute the higher harmonics in the current induced by an {\em ac} bias
of the form given in Eq. \ref{td-bias}. We generally take
$V_\text{dc}= -V_\text{ac}$ in order to ensure a smooth rise of the
bias. This avoids high-frequency components in the current that are
only related to initial transients and also increases the numerical
stability of the time propagation. Both $V_\text{dc}$ and
$V_\text{ac}$ can be independently adjusted in the experiment and
were chosen to be on the same order in recent measurements
\cite{Ward2010,Arielly2011}. Throughout the paper we work with an ac
frequency of $f =$ 200 THz, which correspond to a photon energy
of $\hbar \omega_0 =$ 1.26 eV in the near-infrared. 

\begin{figure}[h]
  \centering
  \includegraphics[width=0.45\textwidth]{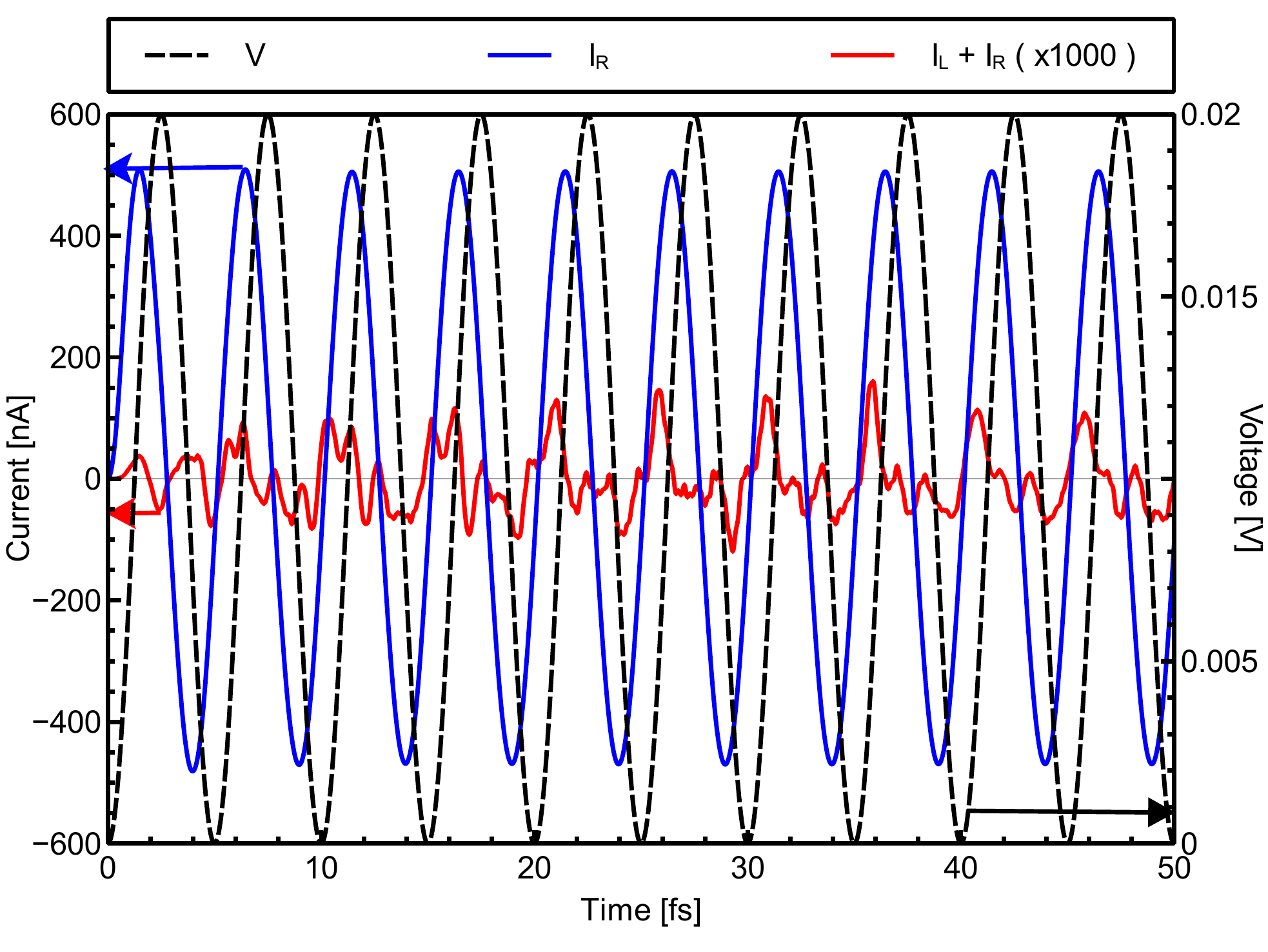}
  \caption{Time dependent bias $V_\text{dc}=
-V_\text{ac}$ = 0.01 V applied to the right lead and
resulting current for the benzenediol
junction from approximate TD-DFT. Shown is also the sum of the current
through the right ($I_R$) and ($I_L$) interface magnified by a factor
of 1000.}    \label{IVlow}
\end{figure}

We first compute the time-dependent current for low bias $V_\text{dc}=
-V_\text{ac}$ = 0.01 V with the method presented in Sec. \ref{theo}. As shown in Fig. \ref{IVlow}, the current leads
the voltage as expected for capacitive transport. Although the bias
polarity always remains positive, the current reverses sign. This
indicates that the conductance $G(\omega_0)$ is larger than the {\em dc}
conductance. This matches with our earlier calculations on the
admittance of this device \cite{Oppenlander2013}. Since the current may be evaluated
at the left and right device boundary, we also depict the sum of
both in  Fig. \ref{IVlow}. In the {\em dc} limit, $I_L$ and $I_R$ 
coincide, while for a time-dependent bias the device region will in
general continuously charge and discharge. The displacement current we
find here is rather small\footnote{For a deeper discussion of
  displacement currents see the introductory article by Oriols and
  Ferry in this special issue \cite{Oriols2013}.}.

\begin{figure}[h]
  \centering
  \includegraphics[width=0.45\textwidth]{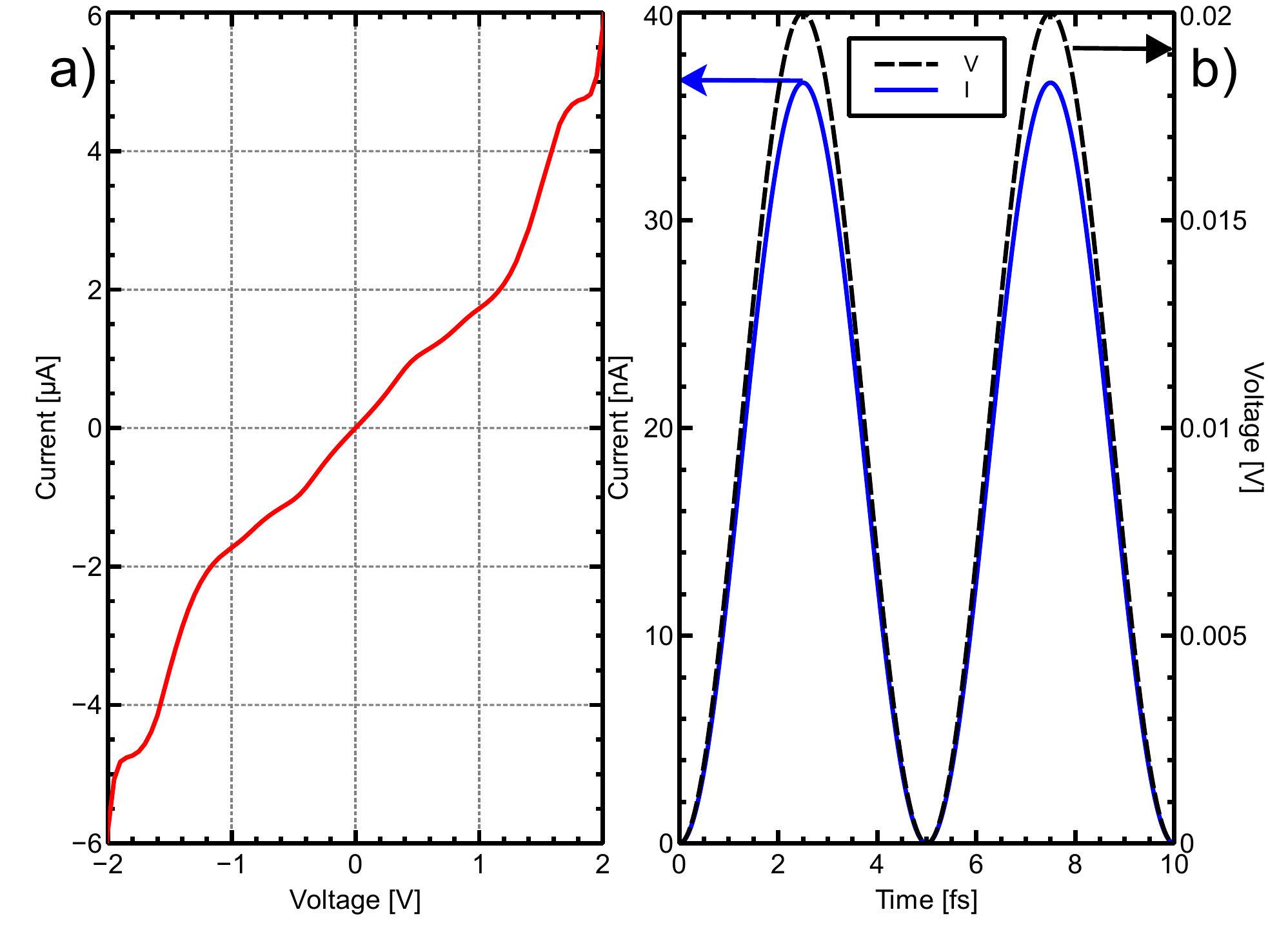}
  \caption{a) Current-voltage characteristic
    $I_\text{dc}(V_\text{dc})$  as obtained from a static Landauer-B\"{u}ttiker
    approach. b) Time-dependent voltage $V(t)$ and quasi-static $I_\text{dc}(V(t))$.}  \label{IVfit}
\end{figure}

In order to examine the accuracy of the quasi-static picture of
Eq. \ref{class} we proceed as follows. First, we compute the
$I_\text{dc}(V_\text{dc})$ curve using the conventional steady-state
Landauer-B\"{u}ttiker formalism based on the DFTB Hamiltonian. As in Sec. \ref{adm}, the same self energy is used in the static and
time-dependent calculations to ensure a common methodological
footing. The I-V curve is sampled in the interval [-2, 2] V with
an increment of 0.05 V and then fitted to a high-order polynomial. The
result is shown in Fig. \ref{IVfit}a). Due to the
symmetric structure with equal lead-molecule distances
on both sides of the device, the I-V characteristic has
inversion symmetry around the origin. For low bias the current rises
roughly linearly, while around 2 V a steep increase is observed, which
arises due to the LUMO\footnote{LUMO = lowest unoccupied molecular
  orbital} of the molecule entering the bias window
\cite{Wang2011}. 

\begin{figure}[h]
  \centering
  \includegraphics[width=0.45\textwidth]{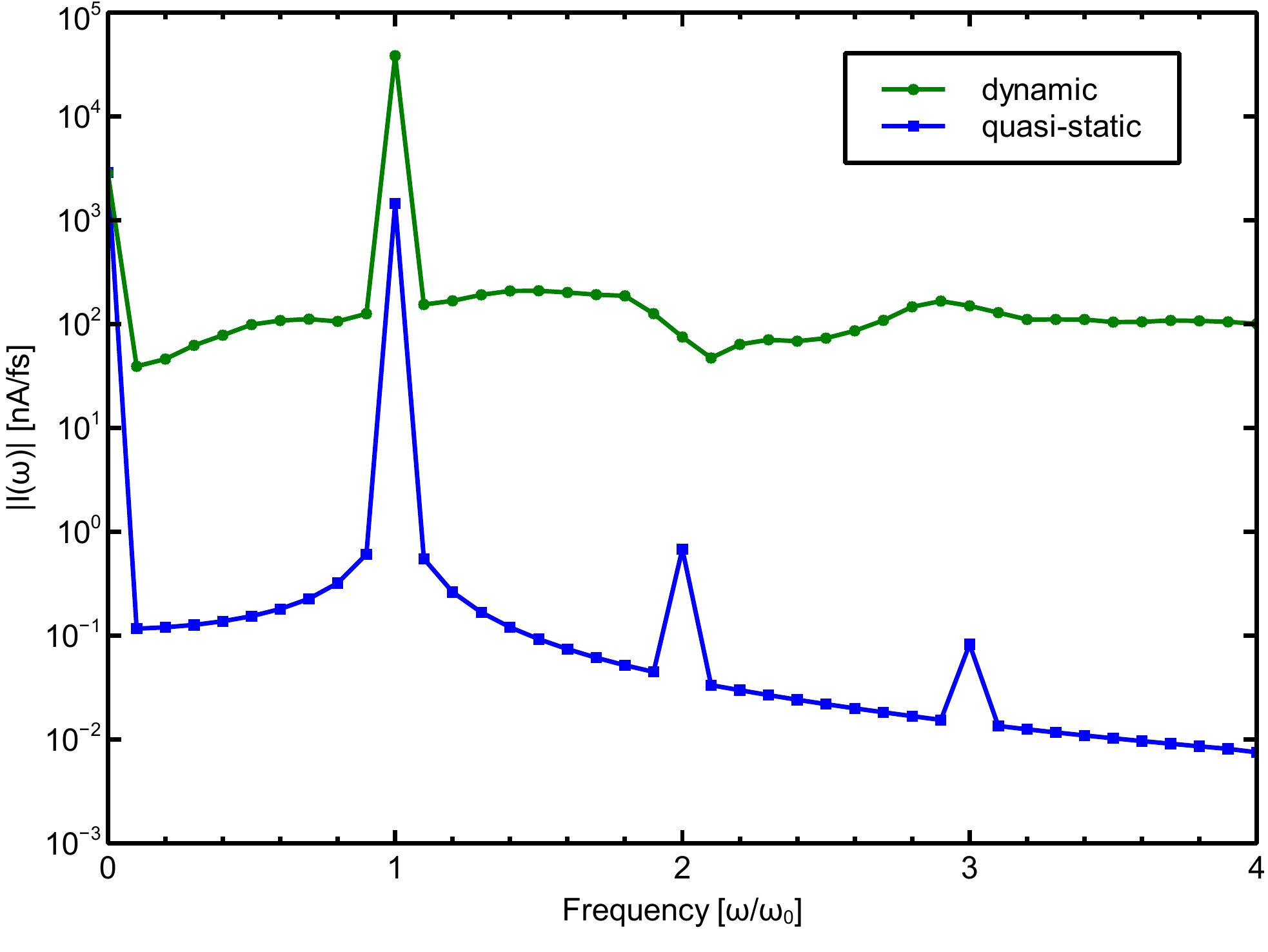}
  \caption{Fourier transform $|I(\omega)|$ of the time-dependent current
    for the dynamic transport simulations and in the quasi-static
    picture. The frequency is given in units of the driving frequency
    $\omega_0$. In both cases the transform was evaluated for a total
    simulation time of 50 fs, resulting in an energy resolution of
    approximately 0.08 eV.  $V_\text{dc}=
-V_\text{ac}$ = 0.01 V.}  \label{FT0.01}
\end{figure}

Evaluating this function $I_\text{dc}(V)$ at the
time-dependent bias Eq. \ref{td-bias}, the quasi-static current in
Fig. \ref{IVfit}b) is obtained. Clearly, current and voltage are in
phase and always of same polarity under this approximation in
disagreement with the result given in Fig. \ref{IVlow}. Further
comparisons can be made by taking the Fourier transform of the current
traces as given in Fig. \ref{FT0.01}. It can be seen that in the
quasi-static approximation a larger number of harmonics exists, while
the dynamic simulation shows a strong peak only at the fundamental
frequency $\omega_0$. As already evident from the polarity of the
current in the time domain, we find $I(\omega_0)>I(\omega=0)$. Since
$I(\omega=0)$ represents the {\em dc} component of the {\em ac} current, or in
other words its time averaged value, we can also easily access the
rectified current from the Fourier transform. Because of the small
non-linearities in this low voltage regime (as also quantified in
Fig. \ref{FT0.01}), we compute in the
quasi-static limit  a value of only 0.8 \% for the ratio
$(\overline{I(t)}-I_\text{dc})/I_\text{dc}$. Although the curvature at $V_\text{dc}$ is
positive (cf. Eq. \ref{sec}), a {\em negative} value of -1.1 \% is
obtained for the dynamic
simulations, showing that quantum effects can be important also for small
dc bias.

\begin{figure}[h]
  \centering
  \includegraphics[width=0.45\textwidth]{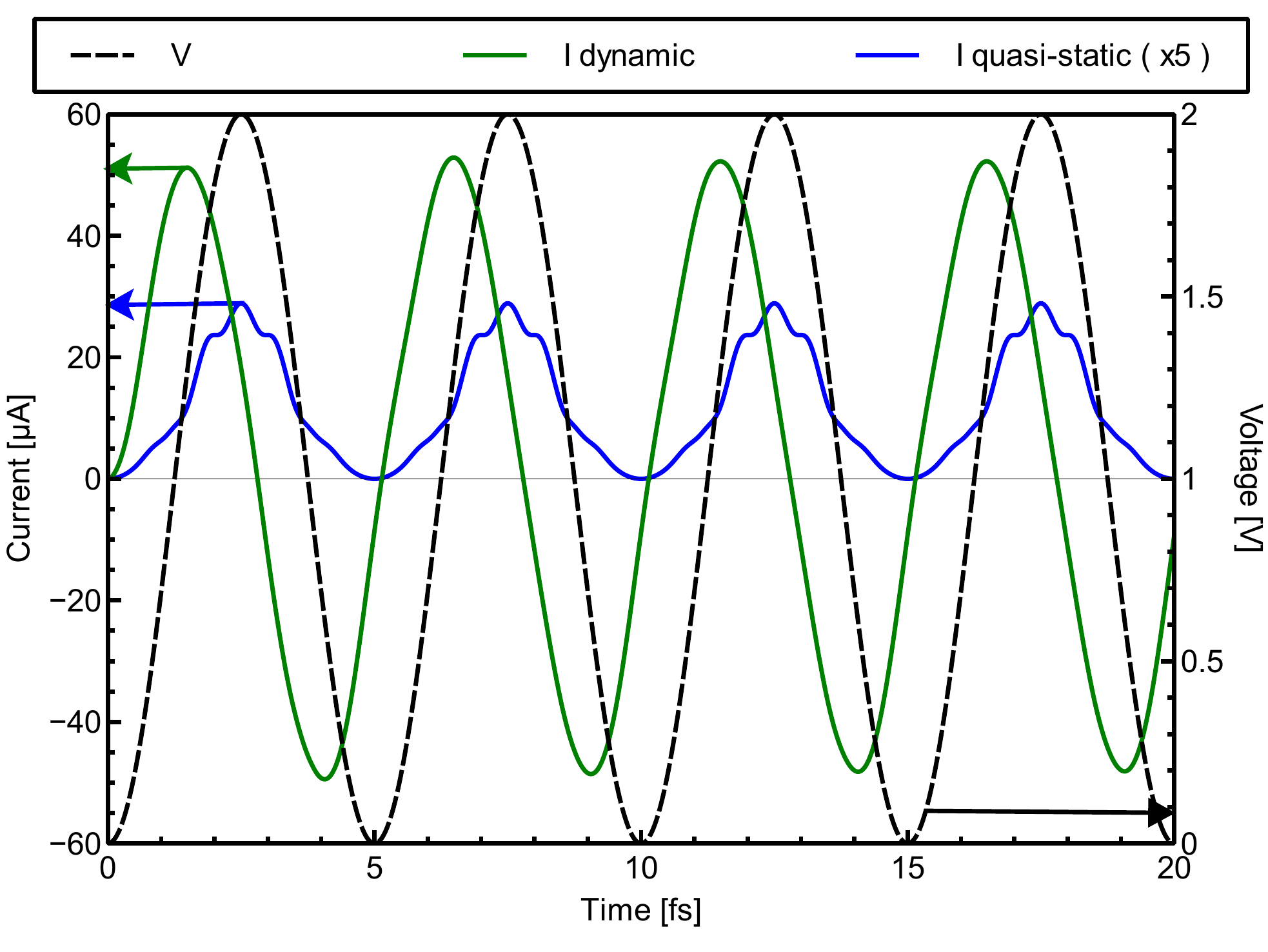}
  \caption{Dynamic and quasi-static currents for $V_\text{dc}=
-V_\text{ac}$ = 1 V. The latter is magnified by a factor of
five. Shown are the first 20 fs out of a total simulation time of 50 fs.}  \label{IV1V}
\end{figure}

\begin{figure}[h]
   \centering
  \includegraphics[width=0.45\textwidth]{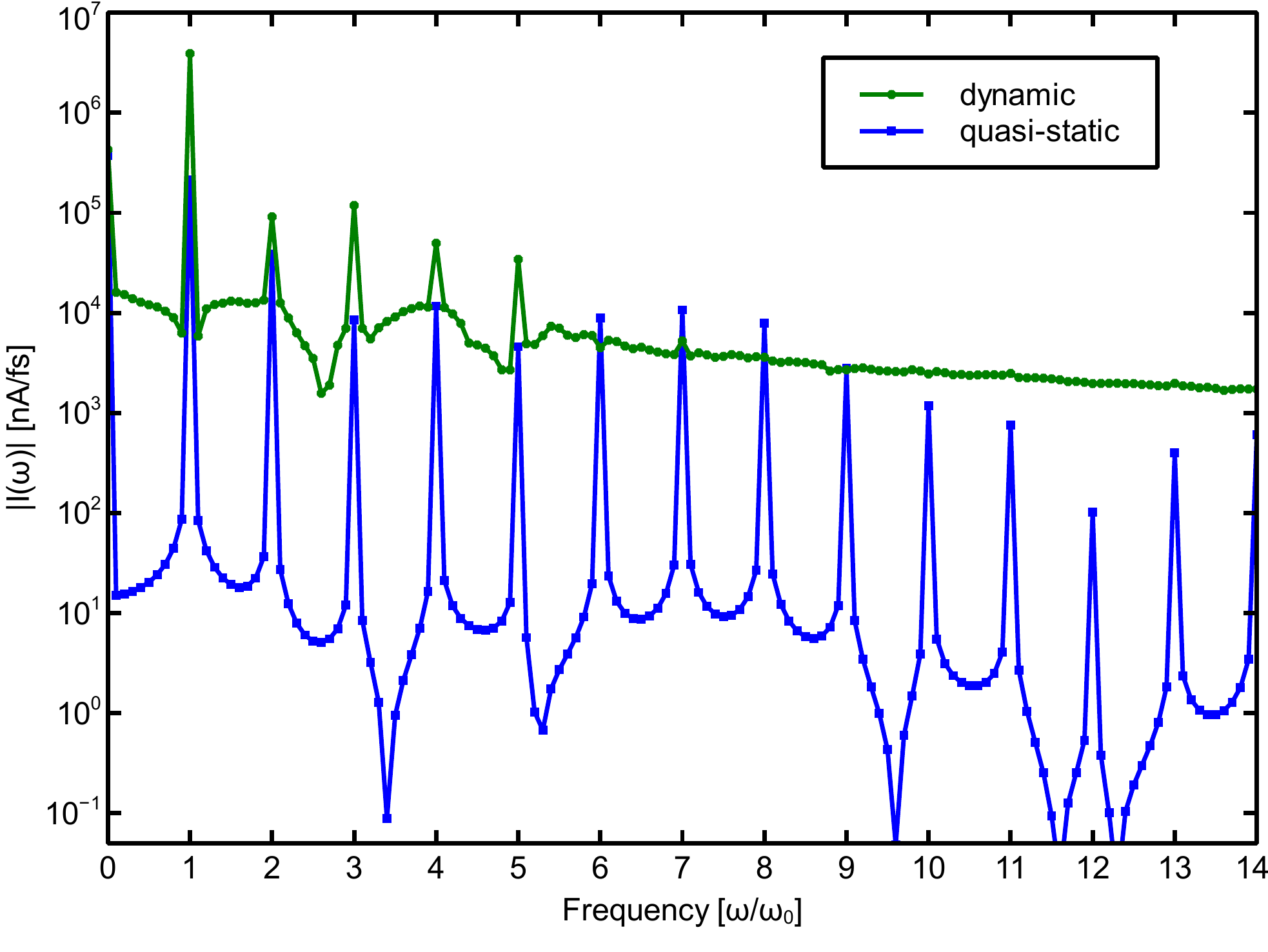}
  \caption{ $|I(\omega)|$ of the time-dependent current
    for dynamic and quasi-static simulations for $V_\text{dc}=
-V_\text{ac}$ = 1 V.} \label{FT1}
\end{figure}

Results for larger bias ($V_\text{dc}=
-V_\text{ac}$ = 1 V) are given for the time domain in Fig. \ref{IV1V}
and in the frequency domain in  Fig. \ref{FT1}. While for the dynamic simulations the profile
of I(t) is relatively smooth even at this
larger bias, the quasi-static data has to follow the highly non-linear
behaviour of $I_\text{dc}(V_\text{dc})$ around 2 V. Consequently, the
corresponding Fourier transform features a larger number of higher
harmonics. In contrast, only up to five harmonics are clearly
discernible in the  dynamic simulations which, however, have a larger
amplitude. The rectified current reflects this with an increase of
60.8 \% and 39.7 \% for the dynamic and quasi-static simulations,
respectively. Another difference between the two approaches is that
the quasi-static $|I(\omega)|$ falls off much stronger inbetween the
harmonics. One possible reason for this could be that initial transients have
not fully died out in the quantum calculations. This point could 
be further investigated by much longer simulations.


\section{Conclusions and summary}
In this article we reported on atomistic calculations of dynamical
charge transport through single molecules. The employed time-domain
approach based on density functional theory allows one to study
arbitrary temporal profiles of the applied bias including in particular
also {\em ac} driving. We find for molecular junctions that the classical quasi-static interpretation of high
harmonic generation is  of limited value already for small bias. If
the {\em ac} frequency exceeds the microwave range, the conductance is
significantly different from its {\em dc} value and the capacitance of the
molecular device has to be taken into account, as discussed in
Sec.~\ref{adm}. As a general trend, we observe that a full quantum
treatment leads to smoother current traces and consequently also to a
much smaller number of higher harmonics. For devices with
a featureless transmission, the rectified current is proportional
to d$^2$I$_\text{dc}$/dV$^2$ both in the classical quasi-static
picture as well as in the quantum case \cite{Tucker1985,Cuevas2010}.
Our results show that there is no such simple
relation for molecular junctions.  Dynamical quantum simulations
should therefore play an important role in the interpretation of experiments on
light driven {\em ac} transport in the future.               

\begin{acknowledgements}
Financial support from the Deutsche Forschungsgemeinschaft (GRK 1570
and SPP 1243) is gratefully acknowledged.
\end{acknowledgements}

\bibliographystyle{spphys}       

\end{document}